\documentstyle[12pt,epsf]{article}
\renewcommand{\baselinestretch}{1.35}
\begin{document}
%
%hep-ph/0107296
\begin{flushright}
  July, 2001 \ \\
  OU-HEP-392 \ \\
\end{flushright}
\begin{center}
\large{A chiral theory of strange sea distributions in the nucleon}
\end{center}
\vspace{0mm}
\begin{center}
M.~Wakamatsu\footnote{Email \ : \ wakamatu@miho.rcnp.osaka-u.ac.jp}
\end{center}
\vspace{-4mm}
\begin{center}
Department of Physics, Faculty of Science, \\
Osaka University, \\
Toyonaka, Osaka 560, JAPAN
\end{center}

\vspace{6mm}
%\begin{flushleft}
%\end{flushleft}

\vspace{1mm}
\begin{center}
\small{{\bf Abstract}}
\end{center}
\vspace{-1mm}
\begin{small}
Theoretical predictions are given for the strange sea distributions
in the nucleon based on the flavor SU(3) chiral quark soliton model,
with emphasis upon the asymmetry of quark and antiquark distributions.
We find that the quark-antiquark asymmetry of the strange sea is much
larger for the longitudinally polarized distribution functions than
for the unpolarized ones.
A preliminary comparison with the CCFR data for the unpolarized
$s$-quark distribution and with the LSS fits of the longitudinally
polarized distribution functions is encouraging.
\end{small}

\vspace{8mm}
An incomparable feature of the chiral quark soliton model (CQSM)
as compared with many other effective models of QCD like the
MIT bag model is that it can give reasonable predictions not
only for the quark distributions but also for the {\it antiquark
distributions} \cite{WK99,WW00A}.
This crucially owes to the field theoretical nature
of the model that enables us to carry out
nonperturbative evaluation of the parton distribution functions with
full inclusion of the {\it vacuum polarization effects} in the
rotating mean field of hedgehog shape \cite{WK99,DPPPW96}.
It was already shown
that, {\it without introducing any adjustable parameter} except for
the initial-energy scale of the $Q^2$-evolution, the CQSM can explain
almost all the qualitatively noticeable features of the recent high-energy
deep-inelastic scattering observables. It naturally explains the
excess of $\bar{d}$-sea over the $\bar{u}$-sea in the
proton \cite{Wakam9298,PPGWW99}.
It also reproduces qualitative behavior of the observed longitudinally
polarized structure functions for the proton, the neutron and the
deuteron \cite{WK99,WW00A}.
The most puzzling observation, i.e. unexpectedly small
quark spin fraction of the nucleon, can also be explained in no need of
large gluon polarization at the low renormalization scale
\cite{WY91,WW00B}.
Finally, the model predicts quite large {\it isospin-asymmetry} also
for the {\it spin-dependent sea-quark distributions}
\cite{WK99,WW00A,DPPPW96,DGPW00}, which we expect will be confirmed by
near future experiments.

Our theoretical analyses so far are based on the flavor SU(2) CQSM
so that no account has been taken of the possibility of
strange quark excitations in the nucleon. However, there have been
several experimental indications that $s \bar{s}$ pairs in the
nucleon are responsible for the numbers of non-trivial
effects \cite{EL01}.
An interesting question here is how large the magnitude of this
admixture is, and/or how large the quark-antiquark asymmetry of the
nucleon strange sea distributions is.
Also interesting is whether we {\it do} expect asymmetry of $s$- and
$\bar{s}$-quarks also for the spin-dependent distributions.

To answer these questions, we here use the CQSM generalized to
flavor SU(3) \cite{WAR92,BDGPPP93}.
To proceed, we first recall some basics of the
SU(2) CQSM. It is specified by the effective lagrangian,
\begin{eqnarray}
   {\cal L}_0 = \bar{\psi} \,(\,i \! \not\!\partial - 
   M e^{\,i \gamma_5 \mbox{\boldmath $\tau$}
   \cdot \mbox{\boldmath $\pi$} (x) / f_\pi \,}
   \,) \psi , \label{modlag}
\end{eqnarray}
which describes the effective quark fields with a dynamically
generated mass $M$, interacting with massless
pions. The nucleon (or $\Delta$) in this model appears as a
rotational state of a symmetry-breaking hedgehog object, which
itself is obtained as a solution of self-consistent Hartree problem
with infinitely many Dirac-sea quarks \cite{DPP88,WY91}.
The theory is not a renormalizable one and it is defined with
some ultraviolet cutoff. In the Pauli-Villars regularization
scheme, which is used throughout the present analysis,
what plays the role of a ultraviolet cutoff is the Pauli-Villars
mass $M_{PV}$ obeying the relation
$\left( N_c M^2 / 4 \pi^2 \right) \ln {\left( M_{PV} / M \right)}^2
=  f_\pi^2$ with $f_\pi$ the pion weak decay constant \cite{DPPPW96}.
Using the value of $M \simeq 375 \,\mbox{MeV}$, which is favored
from the phenomenology of nucleon low energy observables,
this relation fixes the Pauli-Villars mass as
$M_{PV} \simeq 562 \,\mbox{MeV}$.
Since we are to use these values of $M$ and $M_{PV}$, there is
{\it no free parameter} additionally introduced into the calculation of
distribution functions \cite{WW00A}.

Now, the principle dynamical assumption of the SU(3) CQSM is as
follows. The first is the embedding of the SU(2) self-consistent
mean-field (of hedgehog shape) into the SU(3) matrix as
\begin{equation}
 U_0^{\gamma_5} (\mbox{\boldmath $x$}) \ = \ \left(
 \begin{array}{cc}
 e^{i \,\gamma_5 \,\mbox{\boldmath $\tau$} \cdot 
 \hat{\mbox{\boldmath $r$}} \,F(r)} & 0 \\
 0 & 1
 \end{array} \right) \,.
\end{equation}
The next is the semiclassical quantization of the rotational motion in the
SU(3) collective space represented as
\begin{equation}
 U_0^{\gamma_5} (\mbox{\boldmath $x$}, t) \ = \ 
 A(t) \,U_0^{\gamma_5} (\mbox{\boldmath $x$}) \,A^\dagger (t) ,
\end{equation}
with
\begin{equation}
 A(t) \ = \ e^{- \,i \,\Omega \,t}, \ \ \ \ \ \Omega \ = \ 
 \frac{1}{2} \,\Omega_a \,\lambda_a \ \in \ \mbox{SU(3)}.
\end{equation}
The semiclassical quantization of this collective rotation leads to
a systematic method of calculation of any nucleon observables including
the parton distribution functions, which is given as a perturbative
series in the collective angular velocity operator $\Omega$.
(This takes the form of a $1 / N_c$ expansion, since $\Omega$ itself
is a $1 / N_c$ quantity.)  In the present study, all the terms
up to the first order in $\Omega$ are consistently taken into account,
according to the formalism explained in \cite{WK99}.
Since the resultant expressions 
for the quark (and/or antiquark) distribution functions are pretty
lengthy, we decided to show them elsewhere and demonstrate only the main
results here.
Several comments are in order, however. The SU(3) symmetry breaking 
effects arising from the effective mass difference between the strange
and nonstrange quarks (it should be an additional parameter of the model)
can in principle be taken into account by using a perturbation method.
To carry it out for parton distribution functions with full account
of the vacuum polarization effects is quite involved,
however. We therefore leave it to future studies.
(This means that we are still 
continuing parameter-free analyses of the parton distribution functions,
although the magnitude of strange quark mixture under
this approximation should rather be taken as upper limits.)
Secondly, some inconsistency is known to exist between the basic dynamical 
assumption of the SU(3) CQSM and the time-order-keeping collective 
quantization procedure of the rotational motion, although the latter is 
believed to resolve the long-standing $g_A$ problem in the SU(2)
model \cite{WW93,Wakam96}.
Here, we simply follow the symmetry conserving approach advocated in
\cite{PWG99}, which amounts to dropping some theoretically
contradictory terms by hand.

\begin{figure}[htbp] \centering
\epsfxsize=14cm 
\epsfbox{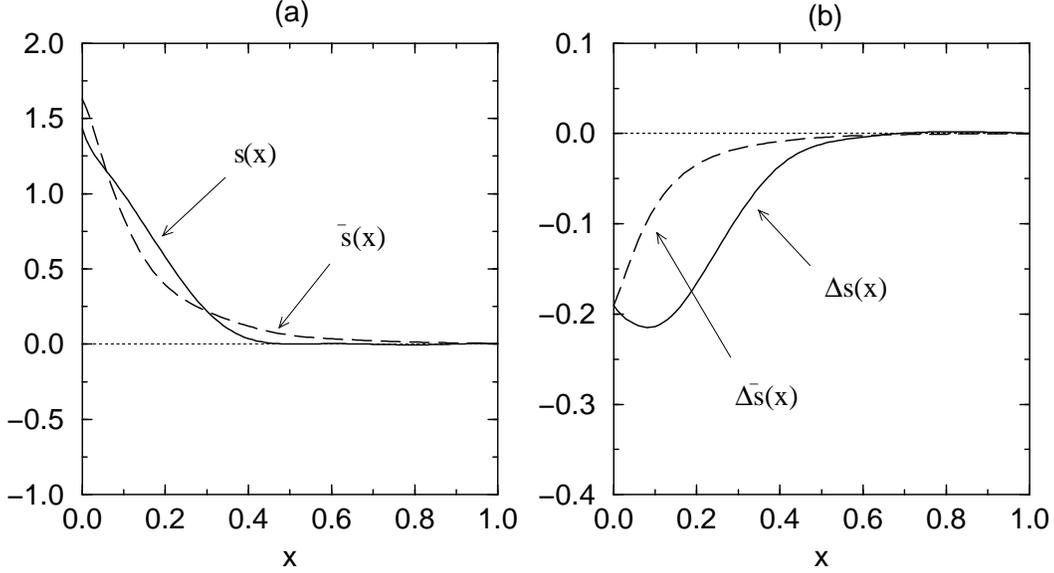}
\renewcommand{\baselinestretch}{1.00}
\caption{The theoretical predictions of the SU(3) CQSM for the
strange sea distributions at the model energy scale.
(a) unpolarized case and (b) longitudinally polarized case.} 
\renewcommand{\baselinestretch}{1.20}
\end{figure}

Now we show in Fig.1 the theoretical $s$- and $\bar{s}$-quark
distribution functions evaluated at the model energy scale.
Here, (a) represents the unpolarized distributions, while (b) does 
the longitudinally polarized distributions.
One confirms that both $s(x)$ and $\bar{s} (x)$ satisfy the
positivity constraint as they should do, in sharp contrast to
the previous result obtained by T\"{u}bingen group in the
so-called ``valence-quark-only'' approximation \cite{SRW99}.
This proves our assertion that the proper account of the vacuum
polarization effects is vital to give any reliable prediction for
anti-quark distributions. One also notices that the
distributions $\bar{s}(x)$ has softer (lower-$x$) component than
$s(x)$ in qualitatively consistent with the argument
of Brodsky and Ma based on the light-cone meson-baryon fluctuation
model \cite{BM96}.
Note, however, that the asymmetry cannot be extremely large due to the 
restriction of the strange-quantum-number conservation in the nucleon,
i.e. $\int_0^1 \,s(x) \,dx = \int_0^1 \,\bar{s} (x) \,dx$ with
$s(x) > 0$ and $\bar{s} (x) > 0$.
In contrast, one observes rather large quark-antiquark asymmetry for
the longitudinally polarized distributions. One sees that the $s$- and 
$\bar{s}$-quarks are both negatively polarized, but $|\Delta \bar{s} (x)|$
is much smaller than $|\Delta s (x)|$.
This feature is again consistent with Brodsky and Ma's conjecture, at least
qualitatively. In fact, they argue that, if the intrinsic strange 
fluctuations in the proton are mainly due to the intermediate
$K^+ \Lambda$ configuration, $s$-quark is negatively polarized but the 
polarization of $\bar{s}$ is zero. Their argument goes as follows.
Since the $K^+$ meson is a pseudoscalar particle with negative parity
and the parity of $\Lambda$ is positive, the parity conservation dictates 
that the relative orbital angular momentum of the intermediate
$K^+ \Lambda$ state must be odd, most probably be a p-wave state.
This gives the total angular momentum wave function in the following
form :
\begin{eqnarray}
 |K^+ \Lambda (J= \frac{1}{2}, J_z = \frac{1}{2}) \rangle
 &=& \sqrt{\frac{2}{3}} \,\,\,| L=1, L_z = 1 \rangle \,| 
 \Lambda (S = \frac{1}{2}, S_z = - \frac{1}{2}) \rangle \nonumber \\  
 &-& \sqrt{\frac{1}{3}} \,\,\,| L = 1, L_z = 0 \rangle \,| 
 \Lambda (S = \frac{1}{2}, S_z = \,\,\,\frac{1}{2}) \rangle \, .
\end{eqnarray}
The point here is that the probability of $\Lambda$-spin being opposite
to the proton spin is twice as large as being parallel to it.
Combining it with the observation that the spin of $\Lambda$ almost
comes from its constituent $s$-quark, one immediately conclude that
the virtually mixed $s$-quark in the proton is negatively polarized
against the proton spin direction.
The situation is quite different for the $\bar{s}$-quark generated 
through the same intrinsic fluctuation $p \rightarrow K^+ \Lambda$.
Since the $\bar{s}$ is contained in the pseudoscalar meson $K^+$ without
spin, the net spin of $\bar{s}$ in the $K^+$ and consequently in the 
proton would be zero. Although qualitatively consistent with this
argument of Brodsky and Ma, the CQSM 
predicts sizable amount of negative polarization also for the
$\bar{s}$-quark.
Such nonzero polarization of $\bar{s}$-quark may be obtained by
introducing more complicated virtual process like 
$p \rightarrow K^{*+} \Lambda$. However, the
precise estimation of the size of polarization in meson cloud models
would be quite hard, since there are many competing processes.

Just by considering intermediate $p \pi^0$ and $n \pi^+$ configurations
instead of $K^+ \Lambda$ fluctuation, the meson-baryon fluctuation model 
(or the meson cloud convolution model) can naturally explain the excess
of $\bar{d}$-sea over the $\bar{u}$-sea in the proton \cite{Kumano98}.
Note, however, that by the same reason as the net polarization of 
$\bar{s}$ is zero, one must conclude that the net polarizations of 
$\bar{d}$- and $\bar{u}$-seas are zero (or at least very small).
This clearly contradicts the previously-mentioned predictions of 
the SU(2) CQSM in which $\Delta \bar{u}(x)$ is large and positive,
while $\Delta \bar{d} (x)$ is large and negative \cite{WW00A,DGPW00}.
In our opinion, what
is responsible for this remarkable difference is the nontrivial
correlation between spin and isospin quantum numbers embedded 
in the CQSM.
At least, one should recognize that the physical contents of the
pion cloud model and the CQSM are not necessarily the same, as naively 
expected. 

There are two popular ways to extract unpolarized strange sea
distributions from the deep-inelastic-scattering data.
The first method uses the 
neutrino-induced charm production, while the second relies upon a global 
fit (like the other flavor densities).
The first direct determination of the strange quark distribution based on
the neutrino-induced charm productions was carried out by the CCFR
collaboration some years ago \cite{CCFR95}.
Here, we perform a very preliminary comparison of the
theoretical predictions of the SU(3) CQSM with the strange quark
distribution obtained by the CCFR next-to-leading-order (NLO) analysis.
The comparison should be taken as preliminary, since the 
hidden strangeness excitation in the nucleon is thought to be very 
sensitive to the inclusion of the SU(3) breaking effects due to the mass 
difference between the strange and nonstrange quarks which we have not yet 
included \cite{SRW99}.
To carry out the comparison, we have taken account of the scale
dependence of the distribution functions by using the Fortran code
of NLO evolution provided by Saga group \cite{HKM98}.
The initial energy scale of this evolution is taken to be 
$Q_{ini}^2 = 0.25 GeV^2$ and the gluon distribution at this scale is 
simply set to be zero, although may not be completely justified.
In Fig.2, we show the theoretical distributions $s(x)$ and $\bar{s}(x)$
together with the result of the CCFR NLO analysis at $Q^2 = 4 GeV^2$
with the constraint $s(x) = \bar{s} (x)$.
Considering that yet-to-be-included SU(3) breaking effects
is expected to suppress the
magnitude of strange quark excitations, it can be said that the theory 
reproduces the order of magnitude of the observed strange sea
distribution. 

\begin{figure}[htbp] \centering
\epsfxsize=7cm 
\epsfbox{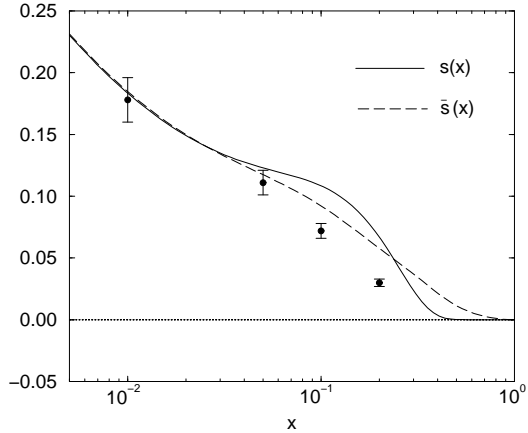}
\renewcommand{\baselinestretch}{1.00}
\caption{The theoretical unpolarized distribution functions
$s (x)$ and $\bar{s} (x)$ at $Q^2 = 4.0 \,\mbox{GeV}^2$
in comparison with the corresponding CCFR data.}
\renewcommand{\baselinestretch}{1.20}
\end{figure}

Turning to the spin-dependent distribution functions, the quality of 
the presently-available semi-inclusive data is rather poor, so that
the main source
of analyses is limited to the inclusive DIS data alone, which forces 
to introduce several simplifying assumption in the analyses.
For instance, most analyses in the past adopt apparently groundless 
assumption of flavor-symmetric polarized sea,
$\Delta \bar{u} (x) =  \Delta \bar{d} (x) = \Delta \bar{s} (x)$
\cite{GS96,GRSV96}.
The other analyses assumes that $\Delta q_3 (x, Q^2) = c \,
\Delta q_8 (x, Q^2)$ with $c$ being a constant. Probably, the most
ambitious analyses free from these {\it ad hoc} assumptions on the
distribution functions are those by Leader, Sidrov and
Stamenov (LSS) \cite{LSS00}. (See also \cite{GRSV01}.)
They also investigated the sensitivity of their analysis 
on the size of SU(3) symmetry breaking effect.
(Although they did not take account of the possibility
$\Delta s(x) \neq \Delta \bar{s} (x)$, it is harmless because only the 
combination $\Delta s(x) + \Delta \bar{s} (x)$ appears in their analysis of
inclusive DIS data.)

To compare the theoretical distributions of the SU(3) CQSM
with the LSS fits given at $Q^2 = 1 GeV^2$, a care must be paid to the
fact that their analyses is carried out in the so-called JET scheme
(or the chirally invariant scheme \cite{Cheng98}).
To take account of it, we start with the theoretical distribution functions
$\Delta u(x) , \Delta \bar{u} (x), \Delta d (x), \Delta \bar{d} (x), 
\Delta s (x), \Delta \bar{s} (x)$, which are taken as initial distribution 
functions given at $Q_{ini}^2 = 0.25 GeV^2$.
Under the assumption that $\Delta g(x) = 0$ at this initial energy scale,
we solve the DGLAP equation in the standard $\overline{MS}$ scheme to
obtain the distributions at $Q^2 = 1 GeV^2$.
The corresponding distribution functions in the JET scheme are
then obtained by using the following transformation :
\begin{eqnarray}
 \Delta \Sigma (x,Q^2)_{JET} &=& 
 \Delta \Sigma (x,Q^2)_{\overline{MS}}
 \ + \ \frac{\alpha_S (Q^2)}{\pi} \,\,N_f \,\,
 (1 - x) \otimes \Delta g(x, Q^2)_{\overline{MS}} \, , \\
 \Delta g(x,Q^2)_{JET} &=& \Delta g(x, Q^2)_{\overline{MS}} \, ,
\end{eqnarray}
with $\Delta \Sigma (x,Q^2) = \sum_{i = 1}^{N_f} 
(\Delta q_i (x, Q^2) + \Delta \bar{q_i} (x, Q^2))$.    

\begin{figure}[htbp] \centering
\epsfxsize=14cm 
\epsfbox{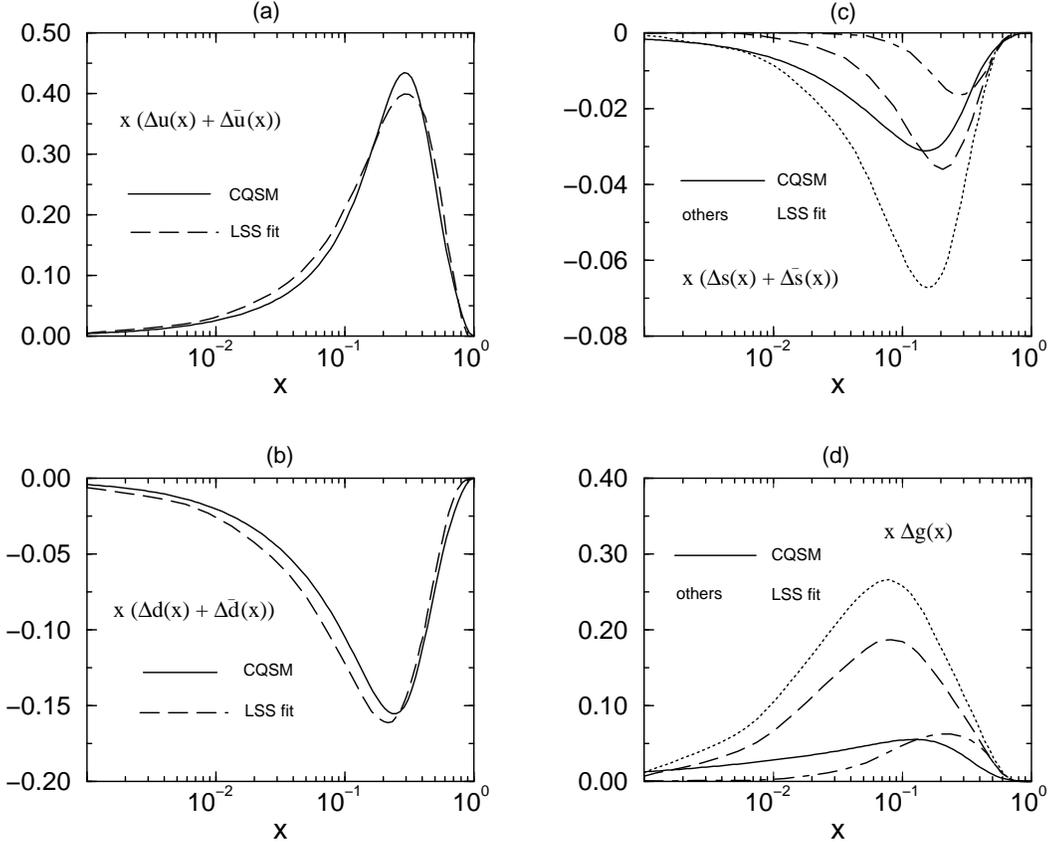}
\renewcommand{\baselinestretch}{1.00}
\caption{The theoretical distribution functions
(a) $x \,(\Delta u(x) + \Delta \bar{u} (x))$,
(b) $x \,(\Delta d(x) + \Delta \bar{d} (x))$,
(c) $x \,(\Delta s(x) + \Delta \bar{s} (x))$, and
(d) $x \,\Delta g(x)$ at $Q^2 = 1.0 \,\mbox{GeV}^2$
in comparison with the corresponding LSS fits in the JET scheme.}
\renewcommand{\baselinestretch}{1.20}
\end{figure}

The solid curves in Fig.3 stand for the theoretical distributions 
$x (\Delta u (x) + \Delta \bar{u} (x)),
x ( \Delta d (x) + \Delta \bar{d} (x)),
x ( \Delta s (x) + \Delta \bar{s} (x))$
and $x \Delta g (x)$ at $Q^2 = 1 GeV^2$ in comparison with the 
corresponding LSS fits.
The long-dashed, dotted and dash-dotted curves in (c) and (d) are 
their fits, respectively obtained by imposing a constraint on the
value of the axial charge $a_8$ to be 0.58 (SU(3) limit), 0.86 and
0.40, while only the case of $a_8 = 0.58$ is shown in (a) and (b)
since these distributions are insensitive to the variation of $a_8$. 
One sees that the distributions $\Delta s(x) + \Delta \bar{s} (x)$
as well as $\Delta g(x)$ are fairly sensitive to the effects of 
SU(3) symmetry breaking and their magnitude cannot be determined 
with good precision from inclusive DIS data alone. 
Bearing in mind this large uncertainties in the 
magnitudes of $x (\Delta s (x) + \Delta \bar{s} (x))$ and
$x \Delta g (x)$, the predictions of the SU(3) CQSM are qualitatively 
consistent with the results of LSS analyses.

\begin{figure}[htbp] \centering
\epsfxsize=14cm 
\epsfbox{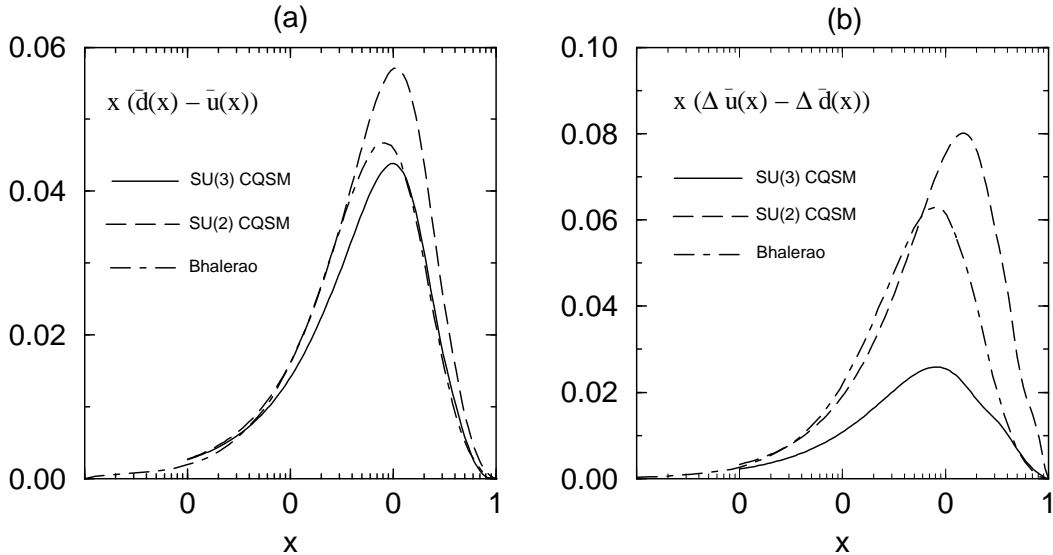}
\renewcommand{\baselinestretch}{1.00}
\caption{The theoretical predictions of the SU(2) and SU(3) CQSM
for (a) $x \,(\bar{d}(x) - \bar{u}(x))$ and
(b) $x \,(\Delta \bar{u}(x) - \Delta \bar{d}(x))$ at
$Q^2 = 0.88 \,\mbox{GeV}^2$ in comparison with
Bhalerao's semi-theoretical predictions.}
\renewcommand{\baselinestretch}{1.20}
\end{figure}

As already emphasized, a noteworthy feature of the SU(2) CQSM is that it
predicts quite large violation of SU(2) symmetry not only for the 
spin-independent sea-quark distributions but also for longitudinally
polarized one.
Why is this observation so important?
This is because the NMC observation $\bar{d} (x) - \bar{u} (x) > 0$
in the proton can be explained equally well by the CQSM and the naive
meson cloud convolution model, whereas the latter essentially predicts
$\Delta \bar{u} (x) \simeq \Delta \bar{d} (x) \simeq 0$ in contrast 
to the prediction of the CQSM such that $\Delta \bar{u} (x) > 0 > 
\Delta \bar{d} (x), | \Delta \bar{u} (x) - \Delta \bar{d} (x)| \simeq 
| \bar{u} (x) - \bar{d} (x) |$.
Now the question is whether this feature of the SU(2) CQSM is also 
shared by the SU(3) CQSM. 
In Fig.4, we compare the predictions of the SU(2) and SU(3) CQSM for
$x (\bar{d} (x) - \bar{u} (x))$ and $x (\Delta \bar{u} (x) -
\Delta \bar{d} (x))$ with the corresponding predictions given
by Bhalerao based on what-he-call the statistical
model \cite{Bhalerao01}. Note that his model is a 
semi-phenomenological one which uses several experimental information
as inputs. One sees that both predictions of the SU(2) and SU(3) CQSM
for $x (\bar{d} (x) - \bar{u} (x))$ are fairly close to 
Bhalerao's prediction which reproduces the NMC data. On the other hand, 
the magnitude of $x (\Delta \bar{u} (x) - \Delta \bar{d} (x))$ is much 
smaller in the SU(3) model than in the SU(2) model. 
This reduction is due to a delicate cancellation of several terms 
in this more complicated theory.
We however conjecture that the introduction of the SU(3) symmetry breaking 
effect in the latter model partially pull back its prediction for 
$x (\Delta \bar{u} (x) - \Delta \bar{d} (x))$ toward that of 
the SU(2) model, thereby leading to the result, which is not extremely
far from Bhalerao's prediction.

In summary, we have given {\it no-free-parameter} theoretical
predictions for the strange sea distributions in the nucleon on
the basis of the flavor SU(3) CQSM. It has been shown that the
$s$- and $\bar{s}$-quarks are both negatively polarized but the
magnitude of $\Delta s(x)$ is much larger than that of
$\Delta \bar{s} (x)$, while the quark-antiquark asymmetry of the
unpolarized strange sea is not extremely large because of the
strange-quantum-number conservation in the nucleon. A preliminary
comparison with the CCFR data for the unpolarized $s$-quark
distribution is encouraging. The theory also reproduces the
characteristic features of the recent LSS fits of the longitudinally
polarized distribution functions including the negatively polarized
strange sea.
We also emphasize that the SU(2) symmetry of the polarized
nonstrange seas is likely to be significantly violated such that
$\Delta \bar{u} (x) > 0 > \Delta \bar{d} (x)$. At any rate,
the {\it spin and flavor dependence} of the
{\it antiquark distributions}
in the nucleon seems very sensitive observables to the nonperturbative
dynamics of QCD at low energy. To reveal this interesting aspect of
QCD, it is vital to carry out various types of high-energy
DIS experiments, which enables us to perform {\it flavor} and
{\it valence plus sea} quark decompositions of the parton
distribution functions.

\vspace{10mm}
\noindent
\begin{Large}
{\bf Acknowledgement}
\end{Large}
\vspace{3mm}

This work is supported in part by the Monbu-kagaku-sho Grant-in-Aid
for Scientific Research No. C-12640267.

%
%  Reference
%

\setlength{\baselineskip}{5mm}


\begin{thebibliography}{99}
\bibitem{WK99}  M.~Wakamatsu and T.~Kubota, Phys. Rev. D {\bf 60},
034020 (1999).
\bibitem{WW00A} M.~Wakamatsu and T.~Watabe, Phys. Rev. D {\bf 62},
054009 (2000).
\bibitem{DPPPW96} D.I.~Diakonov, V.Yu.~Petrov, P.V.~Pobylitsa,
M.V.~Polyakov, and C.~Weiss, \\
Nucl. Phys. B {\bf 480}, 341 (1996) ;
{\it ibid.}, Phys. Rev. D {\bf 56}, 4069 (1997).
\bibitem{Wakam9298} M.~Wakamatsu, Phys. Rev. D {\bf 46}, 3762 (1992) ;\\
M.~Wakamatsu and T.~Kubota, Phys. Rev. D {\bf 57}, 5755 (1998).
\bibitem{PPGWW99} P.V.~Pobylitsa, M.V.~Polyakov, K.~Goeke, T.~Watabe
and C.~Weiss, \\
Phys. Rev. D {\bf 59}, 034024 (1999).
\bibitem{WY91} M.~Wakamatsu and H.~Yoshiki, Nucl. Phys. A {\bf 524},
561 (1991).
\bibitem{WW00B} M.~Wakamatsu and T.~Watabe, Phys. Rev. D {\bf 62},
054009 (2000).   
\bibitem{DGPW00} B.~Dressler, K.~Goeke, M.V.~Polyakov, and C.~Weiss,
Eur. Phys. J. C {\bf 14}, 147 (2000).
\bibitem{EL01} J.R.~Ellis, Nucl. Phys. A {\bf 684}, 53 (2001).
\bibitem{WAR92} H.~Weigel, R.~Alkofer, and H.~Reinhardt,
Nucl. Phys. B {\bf 378} 638 (1992).
\bibitem{BDGPPP93} A.~Blotz, D.~Diakonov, K.~Goeke, N.W.~Park,
V. Yu~Petrov, and P.V.~Pobylitsa, \\ Nucl. Phys. A {\bf 555}, 765
(1993).
\bibitem{DPP88} D.I.~Diakonov, V.Yu.~Petrov, and P.V.~Pobylitsa,
Nucl. Phys. B {\bf 306}, 809 (1988).
\bibitem{WW93} M.~Wakamatsu and T.~Watabe, Phys. Lett. B {\bf 312},
184 (1993) ;\\
Chr.V.~Christov, A.~Blotz, K.~Goeke, P.~Pobylitsa,
V.Yu.~Petrov, M.~Wakamatsu, \\
and T.~Watabe, Phys. Lett. B {\bf 325}, 467 (1994).
\bibitem{Wakam96} M.~Wakamatsu, Prog. Theor. Phys. {\bf 95}, 143
(1996).
\bibitem{PWG99} M.~Prasza{\l}owicz, T.~Watabe, and K.~Goeke,
Nucl. Phys. A {\bf 647}, 49 (1999).
\bibitem{SRW99} O.~Schr\"{o}der, H.~Reinhardt, and H.~Weigel,
Nucl. Phys. A {\bf 651}, 174 (1999).
\bibitem{BM96} S.J.~Brodsky and B.-Q.~Ma, Phys. Lett. B {\bf 381},
317 (1996).  
\bibitem{Kumano98} S.~Kumano, Phys. Rep. {\bf 303}, 183 (1998).
\bibitem{CCFR95} CCFR Collaboration, A.O.~Bazarko et al.,
Z. Phys. C {\bf 65}, 189 (1995).
\bibitem{HKM98} M.~Hirai, S.~Kumano, and M.~Miyama, Compt.
Phys. Commun. {\bf 108}, 38 (1998) ; \\
{\it ibid.}, {\bf 111}, 150 (1998).
\bibitem{GS96} T.~Gehrmann and W.J.~Stirling, Phys. Rev. D {\bf 53},
6100 (1996).
\bibitem{GRSV96} M.~Gl\"{u}ck, E.~Reya, M.~Stratmann, and
W.~Vogelsang, Phys. Rev. D {\bf 53}, 4775 (1996).
\bibitem{LSS00} E.~Leader, A.V.~Sidorov, D.B.~Stamenov, Phys. Lett.
B {\bf 488}, 283 (2000).
\bibitem{GRSV01} M.~Gl\"{u}ck, E.~Reya, M.~Stratmann, and
W.~Vogelsang, Phys. Rev. D {\bf 63}, 094005 (2001).
\bibitem{Cheng98} H.-Y.~Cheng, Phys. Lett. B {\bf 427}, 371 (1998).
\bibitem{Bhalerao01} R.S.~Bhalerao, Phys. Rev. C {\bf 63}, 025208
(2001).
\end{thebibliography}
\end{document}